# Spin-dependent shot noise of inelastic transport through molecular quantum dots

Kamil Walczak

Institute of Physics, Adam Mickiewicz University
Umultowska 85, 61-614 Poznań, Poland

Here we present a theoretical analysis of the effect of inelastic electron scattering on spin-dependent transport characteristics (conductance, current-voltage dependence, magnetoresistance, shot noise spectrum, Fano factor) for magnetic nanojunction. Such device is composed of molecular quantum dot (with discrete energy levels) connected to ferromagnetic electrodes (treated within the wide-band approximation), where molecular vibrations are modeled as dispersionless phonons. Non-perturbative computational scheme, used in this work, is based on the Green's function theory within the framework of mapping technique (GFT-MT) which transforms the many-body electron-phonon interaction problem into a single-electron multi-channel scattering problem. The consequence of the localized electron-phonon coupling is polaron formation. It is shown that polaron shift and additional peaks in the transmission function completely change the shape of considered transport characteristics.

Key words:  shot noise, inelastic transport, molecular quantum dot, electron-phonon interaction, molecular electronics

PACS numbers:  73.38.-k, 73.40.Gk, 73.63.Kv, 85.65.+h

## 1. Introduction

Recent progress in molecular electronics has made it possible to fabricate and to study transport properties of molecular devices [1-9]. Such devices are composed of single molecules (or molecular layers) connected to two (or more) electrodes. A molecule itself represents quantum dot with discrete energy levels, at least an order of magnitude smaller than semiconductor quantum dots (SQD). Since molecule-metal contact is sufficiently weak, molecular quantum dot (MQD) is electrically isolated from metallic electrodes via potential barriers [10-12]. However, in contrast to rigid SQD, molecules involved into the conduction process can be thermally activated to vibrations at finite temperatures. The electrons passing through energetically accessible molecular states (conducting channels) may exchange a definite amount of energy with the nuclear degrees of freedom, resulting in an inelastic component to the current. Such molecular oscillations can have essential influence on the shape of transport characteristics especially in the case, when the residence time of a tunneling electron on a molecular bridge is of order of magnitude of the time involved in nuclear vibrations ( $\sim ps$ ).

Inelastic tunneling across thin films was observed a long time ago, where the particular peaks in the conductance spectra have occurred at various characteristic voltages corresponding to vibrational frequencies of molecules contained in the junction [13]. Inelastic electron tunneling spectroscopy (IETS) performed with scanning tunneling microscope (STM apparatus) was used in later measurements of the conductance of metallic surfaces covered by adsorbates [14-17] or single molecules adsorbed on metallic substrates [18-21]. Recently, the fabrication method of metal single electron transistors (SETs) on scanning tips was used in order to obtain small molecular junction based on different conjugated molecules and conductance spectra for different temperatures, source-drain, and gate voltages were studied [21]. Such experiments give structural information on the molecular junction and provide a direct access to the dynamics of energy relaxation (and consequently thermal dissipation during the tunneling process). It is step further in order to understand the electron conduction at the molecular scale and to model transport characteristics of molecular junctions correctly.

Existing calculations of transport in molecular devices have mostly been focused on current-voltage (I-V) dependences. In particular, it is well known that the electrical current is strongly affected by: (i) the electronic structure of the molecule, (ii) the strength of the coupling with the electrodes, (iii) the location of Fermi level in relation to particular energy levels of the molecule and (iv) the voltage drop along the molecular bridge under applied bias. However, the current itself is not enough to fully characterize the transport, since the current fluctuations (like shot noise) provide additional information regarding to the effective charge of the carriers and their statistics [22-27]. Shot noise is a direct consequence of charge quantization and is unavoidable even at zero temperature as the only source of noise.

When molecule is bridging ferromagnetic electrodes, all the transport characteristics are spin-dependent, where the magnitude of the current flowing through the device and its fluctuations depend on the relative orientation of magnetizations in the electrodes [28-32]. In particular, spin-polarized transport of electrons tunneling through the junction consisting of a self-assembled monolayer (SAM) of octanethiol attached to a pair of Ni electrodes was studied experimentally [33]. These molecular junctions exhibit magnetoresistance values up to 16 % at low bias voltages. However, strong voltage and temperature dependence of the junction magnetoresistance and time-dependent telegraph noise signals suggest that transport properties of the mentioned device can be affected by localized states in the molecular monolayer. The main purpose of this work is to study the influence of molecular vibrations on transport characteristics of magnetic nanodevices. Anyway, noise measurements in magnetic molecular-scale junctions still remain a certain challenge in molecular transport.

## 2. Theoretical background

2.1 Description of the model and mapping technique.

Let us consider MQD represented by discrete energy levels, which is weakly connected with two electrodes through tunnel barriers. Furthermore, let us assume that the electrons occupying the dot interact locally with optical (dispersionless) phonon excitations (the presence of phonons is restricted to the molecular region). To carry out our analysis we can write the full Hamiltonian of considered system as

$$H = \sum_\alpha H_\alpha + H_M + H_T,  \quad (1)$$

where $\alpha = L$ for the left electrode and $\alpha = R$ for right one, respectively. Both metallic electrodes are treated as reservoirs of non-interacting electrons and described with the help of the following Hamiltonian

$$H_\alpha = \sum_{k,\sigma \in \alpha} \varepsilon_k c_{k\sigma}^+ c_{k\sigma}.  \quad (2)$$

Here $\varepsilon_k$ is the single particle energy of conduction electrons, while $c_{k\sigma}^+$ and $c_{k\sigma}$ denote the electron creation and annihilation operators with momentum $k$ and spin $\sigma$. The third term in Eq.1 represents molecular Hamiltonian written in the Holstein-type form

$$H_M = \sum_{i,j}\left[\varepsilon_i - \lambda_j(a_j + a_j^+)\right]d_i^+ d_i + \sum_j \omega_j a_j^+ a_j . \qquad (3)$$

Here $\varepsilon_i$ is single energy level of the molecule, $\omega_j$ is phonon energy in mode $j$, $\lambda_j$ is the strength of on-level electron-phonon interaction. Furthermore, $d_i^+$ and $d_i$ are electron creation and annihilation operators on level $i$, while $a_j^+$ and $a_j$ are phonon creation and annihilation operators, respectively. The last term in Eq.1 describes the connection between MQD and two electrodes

$$H_T = \sum_{k,\sigma\in\alpha;i}\left[\gamma_{k\sigma,i} c_{k\sigma}^+ d_i + h.c.\right], \qquad (4)$$

where the matrix elements $\gamma_{k\sigma,i}$ stands for the strength of the tunnel coupling between the dot and ferromagnetic electrodes. To simplify the notation, throughout this work we use atomic units with $e = \hbar = 1$, so current is given in $e/\hbar$, while noise is given in $e^2/\hbar$.

The problem we are facing now is to solve a many-body problem with phonon emission and absorption when the electron tunnels through the dot. Let us consider for transparency only one phonon mode (primary mode), since generalization to multi-phonon mode case can be obtained straightforwardly. The electron states into MQD are expanded onto the direct product states composed of single-electron states and $m$-phonon Fock states

$$|i,m\rangle = d_i^+ \frac{(a^+)^m}{\sqrt{m!}}|0\rangle , \qquad (5)$$

where electron state $|i\rangle$ is accompanied by $m$ phonons ($|0\rangle$ denotes the vacuum state). Similarly the electron states in the electrodes can be expanded onto the states

$$|k\sigma,m\rangle = c_{k\sigma}^+ \frac{(a^+)^m}{\sqrt{m!}}|0\rangle , \qquad (6)$$

where the state $|k\sigma\rangle$ with momentum $k$ and spin $\sigma$ is accompanied by $m$ phonons. In this procedure, the reservoir Hamiltonian (Eq.2) is mapped to a multichannel model

$$\tilde{H}_\alpha = \sum_{k,\sigma\in\alpha;m}(\varepsilon_{k\sigma} + m\omega)|k\sigma,m\rangle\langle k\sigma,m| . \qquad (7)$$

Since the channel index $m$ represents the phonon quanta excited in the system, accessibility of particular conduction channels is determined by a weight factor

$$P_m = [1 - \exp(-\beta\omega)]\exp(-m\beta\omega) , \qquad (8)$$

where Boltzmann distribution function is used to indicate the statistical probability of the phonon number state $|m\rangle$ at finite temperature $\theta$, $\beta = 1/(k_B\theta)$ and $k_B$ is Boltzmann constant. To determine the temperature of the molecule we assume that the dot is in thermal equilibrium with ferromagnetic electrodes, even under nonequilibrium transport conditions. Thus, here we neglect nonequilibrium phonon effects (due to the assumed high energy relaxation rate) as well as dissipative processes (due to the assumed isolation from the

influence of external surrounding). It means that the electron energies are constrained by the following energy conservation law

$$\varepsilon_{in} + m\omega = \varepsilon_{out} + n\omega. \qquad (9)$$

It is also important to point out that also the tunneling electrons can induce vibrations of the molecular bridge, but in this work we do not consider this mechanism of phonon excitations, assuming that the phonon distribution function is independent on the electric current flowing through the junction. Moreover, in practice, the basis set is truncated to a finite number of possible excitations $m = m_{max}$ because of the numerical efficiency. The size of the basis set strongly depends on: (i) phonon energy, (ii) temperature of the system under investigation and (iii) the strength of the electron-phonon coupling constant. In the new representation (Eq.5), molecular Hamiltonian (given by Eq.3) can be rewritten in the form

$$\tilde{H}_M = \sum_{i,m}(\varepsilon_i + m\omega)|i,m\rangle\langle i,m| - \sum_{i,m}\lambda\sqrt{m+1}\left(|i,m+1\rangle\langle i,m| + |i,m\rangle\langle i,m+1|\right) \qquad (10)$$

which for each molecular energy level $i$ is analogous to tight-binding model with different site energies and site-to-site hopping integrals (see Fig.1). Finally, the tunneling part can also be rewritten in terms of considered basis set of states as

$$\tilde{H}_T = \sum_{k,\sigma\in\alpha;i,m}\left(\gamma_{k\sigma,i}^{m}|k\sigma,m\rangle\langle i,m| + h.c.\right), \qquad (11)$$

where $\gamma_{k\sigma,i}^{m}$ is the coupling between the $m$th pseudochannel in the electrode and the molecular system, respectively. In this way, the many-body electron-phonon interaction problem is mapped into a multi-channel single-electron scattering problem [27,34-40].

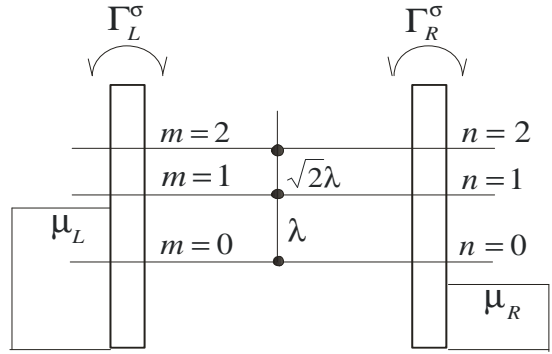

Figure 1: A schematic representation of inelastic scattering problem for the device composed of molecular quantum dot with single energy level connected to the ferromagnetic electrodes.

2.2 Determination of transport characteristics.

To avoid unnecessary complexities, in further analysis we take into account molecular bridge which is represented by one electronic level – generalization to multilevel system is simple. In the Landauer picture, transport through a nanoscopic system is usually described in terms of the transmission probability $T(\varepsilon)$ that a single electron with injection energy $\varepsilon$ scatters from the left electrode – through the molecular bridge – to the right electrode. Molecule itself acts as a strong defect in a periodic structure of two ideal electrodes. When phonon quanta are present on the dot, an electron entering from the left hand side can suffer inelastic collisions by absorbing or emitting phonons before entering the right electrode. Such processes are presented graphically in Fig.1, where individual channels are indexed by the number of phonon quanta in the left $m$ and right electrode $n$, respectively. Each of the possible processes is described by its own transmission probability, which can be written in the factorized form

$$T_{m,n}^{\sigma,\sigma'}(\varepsilon) = \Gamma_L^\sigma \Gamma_R^{\sigma'} \left| G_{m+1,n+1}^{\sigma,\sigma'}(\varepsilon) \right|^2. \qquad (12)$$

Such transmission probability (Eq.12) is expressed in terms of the so-called linewidth functions $\Gamma_\alpha^\sigma$ and the matrix element of the Green's function defined as:

$$G^{\sigma,\sigma'}(\varepsilon) = \left[ 1\varepsilon - \tilde{H}_M - \Sigma_L^\sigma - \Sigma_R^{\sigma'} \right]^{-1}. \qquad (13)$$

Here 1 stands for identity matrix, $\tilde{H}_M$ is the transformed molecular Hamiltonian (Eq.10), while the effect of the electronic coupling to the electrodes is fully described by specifying self-energy corrections $\Sigma_\alpha^\sigma$.

In the present paper we adopt wide-band (WB) approximation to treat ferromagnetic electrodes, where the hopping matrix element is independent of energy, spin and bias voltage, i.e. $\gamma_{k\sigma,i}^m = \gamma_\alpha$. In this case, the self-energy is given through the relation:

$$\Sigma_\alpha^\sigma = -\frac{i}{2}\Gamma_\alpha^\sigma, \qquad (14)$$

where

$$\Gamma_\alpha^\sigma = 2\pi |\gamma_\alpha|^2 \rho_\alpha^\sigma. \qquad (15)$$

Here $\rho_\alpha^\sigma$ is the spin-$\sigma$ band density of states in the $\alpha$-electrode. This self-energy function is mainly responsible for level broadening and generally depends on: (i) the ferromagnetic material that the electrode is made of (Fe, Co, Ni) and (ii) the strength of the coupling with the electrode. There are few factors that can be crucial in determining the parameter of the coupling strength, such as: (i) the atomic-scale contact geometry, (ii) the nature of the molecule-to-electrode coupling (chemisorption or physisorption), (iii) the molecule-to-electrode distance or even (iv) the variation of the surface properties due to adsorption of molecular monolayer. Moreover, both electrodes are also identified with their electrochemical potentials

$$\mu_L = \varepsilon_F - \eta V \qquad (16)$$

and

$$\mu_R = \varepsilon_F + (1-\eta)V, \qquad (17)$$

which are related to the Fermi energy level $\varepsilon_F$ [41]. The voltage division factor $0 \leq \eta \leq 1$ describes how the electrostatic potential difference $V$ is divided between two contacts and can be related to the relative strength of the coupling with two electrodes $\eta = 2^{-\gamma_L/\gamma_R}$. In our analysis we can distinguish two boundary cases: $\gamma_L = \gamma_R \Rightarrow \eta = 1/2$ for interpretation of mechanically controllable break-junction experiments, $\gamma_L \gg \gamma_R \Rightarrow \eta \approx 0$ for interpretation of scanning tunneling microscopy (STM experiments), respectively. Here we assume the symmetric coupling case ($\eta = 1/2$), but it should also be noted that the case of asymmetric coupling ($\eta \neq 1/2$) generates rectification effect [42].

Having transmissions for all the possible transitions (given by Eq.12) we can define the total transmission function as a sum over all the incoming channels $m$ weighted by the appropriate Boltzmann factor $P_m$ and a sum over all the outgoing channels $n$ [35]

$$T_{tot}(\varepsilon) = \sum_{m,n,\sigma} P_m T^{\sigma,\sigma'}_{m,n}(\varepsilon). \tag{18}$$

The elastic part of the transmission (in which the electron preserves its energy) can be achieved by equaling the number of phonons on both electrodes $n = m$, so

$$T_{el}(\varepsilon) = \sum_{m,\sigma} P_m T^{\sigma,\sigma'}_{m,m}(\varepsilon). \tag{19}$$

Transmission function is very important characteristic from the transport viewpoint, since at low voltages the linear conductance $g(=I/V)$ is directly proportional to the convolution of the transmission function $T(\varepsilon)$ and the so-called thermal broadening function $F_T(\varepsilon)$ [41]

$$g(\varepsilon) = \frac{1}{2\pi}\int d\varepsilon' T(\varepsilon') F_T(\varepsilon - \varepsilon'), \tag{20}$$

where

$$F_T(\varepsilon) = \frac{\beta}{4}\sec h^2\left[\beta\frac{\varepsilon}{2}\right]. \tag{21}$$

Since typical thermal energy scale ($1/\beta \sim 0.03 eV$) is relatively small in comparison with transport energy scale ($\sim eV$), we can approximate thermal broadening function with the help of Dirac delta function $F_T(\varepsilon) \approx \delta(\varepsilon)$ and therefore we can identify conductance with transmission $g(\varepsilon) = T(\varepsilon)/2\pi$.

The total current flowing through the junction can be expressed in terms of transmission probability of the particular transitions (given by Eq.12) which connects incoming pseudochannel $m$ with outgoing pseudochannel $n$ [35]

$$I_{tot}(V) = \frac{1}{2\pi}\int_{-\infty}^{+\infty} d\varepsilon \sum_{m,n,\sigma} T^{\sigma,\sigma'}_{m,n}\left[P_m f_L^m\left(1-f_R^n\right) - P_n f_R^n\left(1-f_L^m\right)\right], \tag{22}$$

where

$$f_\alpha^m = \left[\exp[\beta(\varepsilon + m\omega - \mu_\alpha)] + 1\right]^{-1} \tag{23}$$

is the equilibrium Fermi distribution function. The elastic contribution to the current can be obtained from Eq.22 by the assumption of $n = m$, so

$$I_{el}(V) = \frac{1}{2\pi}\int_{-\infty}^{+\infty} d\varepsilon \sum_{m,\sigma} T^{\sigma,\sigma'}_{m,m} P_m \left[f_L^m - f_R^m\right]. \tag{24}$$

Moreover, the magnetoresistance ($MR$) can be defined as a relative difference of the current in the parallel ($P$) and antiparallel ($AP$) configuration of the spin polarization alignment in the electrodes [31,32]

$$MR(V) = \frac{I_P(V) - I_{AP}(V)}{I_{AP}(V)}. \tag{25}$$

The orientation of magnetizations in the electrodes can be changed by applying an external magnetic field.

Shot noise is the time-dependent fluctuation of the electrical current due to the discreteness of the charge of the current carriers and can be computed as the Fourier transform of the current-current correlation function [43]

$$S \equiv \frac{1}{2}\int_{-\infty}^{+\infty} dt \left[\langle I(t)I(0)\rangle + \langle I(0)I(t)\rangle - 2\langle I(t)\rangle\langle I(0)\rangle\right], \quad (26)$$

where $\langle ... \rangle$ represents the quantum statistical average. Limiting ourselves to the final results, the expression for the total spectral density of shot noise in the zero-frequency limit is given through the following relation [27]

$$S_{tot}(V) = \frac{1}{\pi}\int_{-\infty}^{+\infty} d\varepsilon \sum_{m,n,\sigma} \left\{ (T_{m,n}^{\sigma,\sigma'})^2 \left[P_m f_L^m (1-f_L^n) + P_n f_R^n (1-f_R^m)\right] \right. \\ \left. + (1-T_{m,n}^{\sigma,\sigma'})T_{m,n}^{\sigma,\sigma'}\left[P_m f_L^m (1-f_R^n) + P_n f_R^n (1-f_L^m)\right] \right\} \quad (27)$$

The elastic contribution to shot noise can be determined by imposing the constraint of elastic tunneling $m = n$, so

$$S_{el}(V) = \frac{1}{\pi}\int_{-\infty}^{+\infty} d\varepsilon \sum_{m,\sigma} \left\{ T_{m,m}^{\sigma,\sigma'} P_m \left[f_L^m(1-f_R^m) + f_R^m(1-f_L^m)\right] + (T_{m,m}^{\sigma,\sigma'})^2 P_m \left(f_L^m - f_R^m\right)^2 \right\}. \quad (28)$$

The general theory of shot noise in nanoscopic systems allows us to define also the so-called Fano factor with the help of relation [43]

$$F = \frac{S}{2|I|}. \quad (29)$$

Fano factor contains information about electron correlations in the system. Here we can distinguish three different cases: (i) sub-Poissonian shot noise with $F<1$ (electron correlations reduce the level of current fluctuations below unity), (ii) Poissonian shot noise with $F=1$ (there is no correlations among the charge carriers), and (iii) super-Poissonian shot noise with $F>1$ (electron anticorrelations increase the level of current fluctuations above unity).

## 3. Numerical results and their interpretation

In this section we discuss some features of the transport characteristics associated with magnetic single-molecule junction, where molecule bridges ferromagnetic electrodes. By assumption, molecular quantum dot is represented by one electronic level which is coupled to a single vibrational mode. This is a test case simple enough to analyze the essential physics of the problem in detail and control results by manipulating of model parameters. In our calculations we have used the following set of parameters (given in eV): $\varepsilon_i = 0$ (the reference LUMO level), $\varepsilon_F = -1$, $\omega = 1$, $\lambda = 0.5$, $\gamma_L = \gamma_R = 0.2$ (weak coupling is justified by experimental results). Since ferromagnets have unequal spin up and spin down populations, their densities of states for both spin orientations are different. Here we adopt densities of states for Co electrodes from the works [30,31] as obtained from band structure calculations performed using the tight-binding version of the linear muffin-tin orbital method in the atomic sphere approximation: $\rho^\uparrow = 0.1367$, $\rho^\downarrow = 0.5772$ (given in 1/eV). The temperature of the system is set at $\theta = 293$ K ($\beta = 40/eV$). Maximum number of allowed phonon quanta $m_{max} = 4$ is chosen to give fully converged results for all the model parameters.

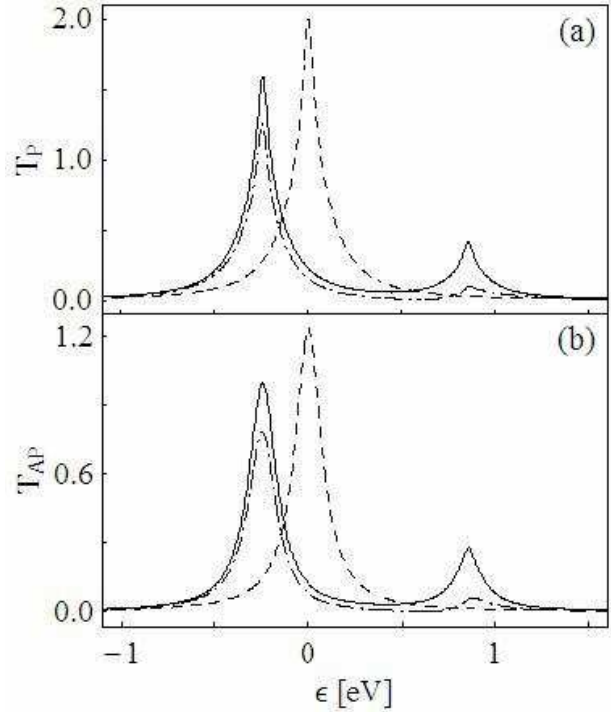

Figure 2: Transmission as a function of electron energy in relation to molecular single level (zero) in the case of parallel (a) and antiparallel alignment of magnetizations (b). Total transmission (solid line) and its elastic part (dashed-dotted line) are compared with transmission probability in the absence of phonons (dashed line).

In Fig.2 we plot energy-dependent transmission functions for analyzed system. When the electron is not coupled to the phonon mode, for one discrete energy level – one resonant transmission peak is observed. However, perfect transmission ($T = 2$ for two channels) is predicted only in the case of $P$ alignment, since for the $AP$ configuration different densities of states are used to calculate particular contributions to transmission function. Generally, the height of a current step is directly proportional to the area of the corresponding transmission peak. It should be also noted that transmission is symmetrical function of energy with respect to resonance ($\varepsilon = 0$). In the presence of electron-phonon coupling, the transmission function reveals additional peaks which indicate the opening of channels involving phonons, while the main peak is reduced in height. Positions of the mentioned peaks approximately coincide with polaron energies, which are given through the relation

$$\varepsilon_{pol}(m) = \varepsilon_i - \frac{\lambda^2}{\omega} + m\omega, \qquad (30)$$

where $m$ denotes the $m$th excited state of a polaron (defined as a state of an electron coupled to phonons). The main peak corresponds to the tunneling through the polaron ground state $\varepsilon_{pol}(0)$, while additional side peaks represent the next excited states with the following energies: $\varepsilon_{pol}(1)$, $\varepsilon_{pol}(2)$, ..., $\varepsilon_{pol}(m_{max})$, respectively. Of course, the separation between the transmission peaks is approximately set by the frequency of the phonon mode $\omega$. Interestingly, all the peaks have elastic and inelastic contributions. Moreover, the localized electron-phonon interaction leads to the so-called polaron shift $\Delta = -\lambda^2/\omega$ which appears as an energy correction for resonant tunneling. There is only one excited polaron state shown in Fig.2, since the intensities of the next excited states are negligibly small (negligible contribution to the current but observed in the logarithmic-scale plot of the transmission function).

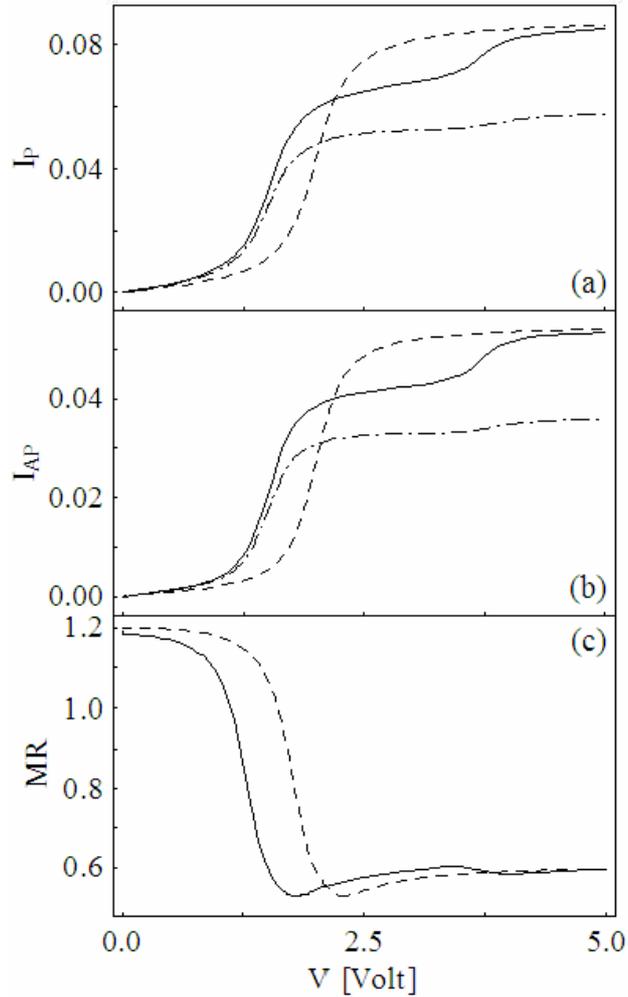

Figure 3: Current-voltage characteristics $I(V) = -I(-V)$ in the case of parallel (a) and antiparallel alignment of magnetizations (b). Total current (solid line) and its elastic part (dashed-dotted line) are compared with the current obtained in the absence of phonons (dashed line). (c) Magnetoresistance $MR(V) = MR(-V)$ as a function of applied bias in the presence (solid line) and absence of phonons (dashed line).

All the features of transmission function discussed earlier are reflected in the tunneling current flowing through the junction, as shown in Figs.3a and 3b. In the absence of phonons, only one step structure occurs when electrochemical potential of the left electrode coincide with the considered LUMO level of MQD. Inclusion of electron-phonon interaction leads to polaron formation and as its consequence to the shift of the main current step, resulting in reduction of the conductance gap (CG). Indeed, some state-of-art first-principles calculations overestimate the CG quantity in comparison with experimental data [44]. Here we indicate that polaron shift can be responsible for such discrepancy. Moreover, the additional resonant peaks in the transmission induce additional current jumps due to emission of phonons. Every step in the I-V dependence has its elastic as well as inelastic contribution. Since $e/\hbar \approx 243.5 \times 10^{-6}$ A, the magnitude of the current flowing through the junction is given in tens of μA, what is comparable with the mentioned *ab initio* results [44]. Anyway, the current for *P* configuration reaches higher values in comparison with the case of *AP* alignment of magnetizations. It should be also noted that in a high-bias limit, the magnitude of inelastic current is always lower and only asymptotically can achieve the values of the current for non-phonon case. In Fig.3c we can see the behavior of magnetoresistance with increasing of bias voltage. In a zero-bias limit $MR \approx 1.2$ for both cases (i.e. in the presence and absence of phonons) and this value immediately decreases when the first current step occurs. Here again polaron formation shifts the step of MR coefficient in the direction to lower voltages. In a high-bias limit, magnetoresistance tends to the same value $MR \approx 0.6$ for both considered cases.

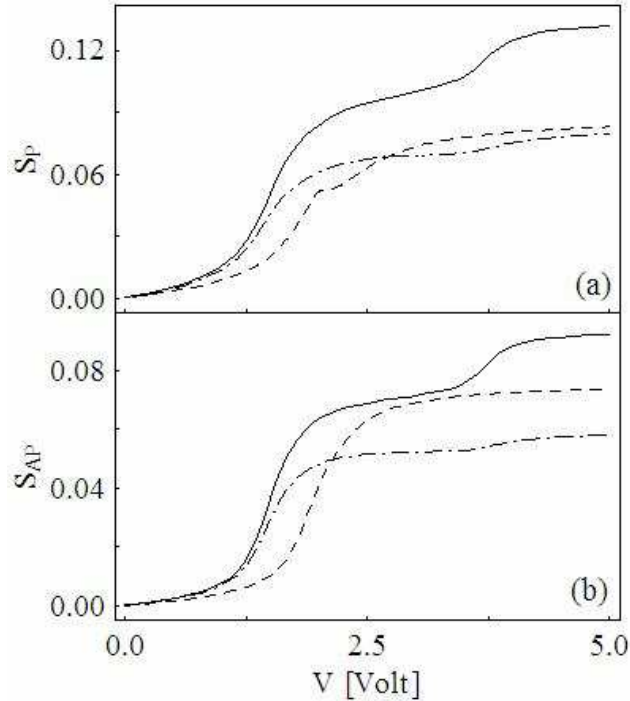

Figure 4: Zero-frequency shot noise $S(V) = S(-V)$ as a function of bias voltage in the case of parallel (a) and antiparallel alignment of magnetizations (b). Total noise power (solid line) and its elastic part (dashed-dotted line) are compared with the noise obtained in the absence of phonons (dashed line).

Obviously, the shape of the shot noise curve is similar to that of the current, as is viewed in Fig.4. The only difference is associated with their behavior in a high bias limit, where inelastic noise power can exceed shot noise for non-phonon case – in opposition to the current-voltage analysis. As was mentioned in the previous section, information about statistical properties of the electrons is included into the Fano factor, which is plotted in Fig.5. Since in our model all the interactions between the current carriers are neglected, such electron correlations are associated only with the Pauli principle. This exclusion rule is related to the fact that one electron feels the presence of the others, since it can not occupy the state already occupied by the electron with the same spin. The crossover in the shot noise power from Poissonian limit ($F=1$) to sub-Poissonian region ($F<1$) is always observed after the first step in the I-V dependence. It means that electrons tunnel in a correlated way. The important thing is significant enhancement of Fano factor due to the phonon effects, observed for $V>2$ Volts, where multi-channel process reduces electron correlations in comparison with single-channel one. Moreover, polaron shift can also be easily recognized in Fig.5. Finally, in a high bias limit, Fano factor for $AP$ alignment of magnetizations reaches bigger values than in the case of $P$ configuration.

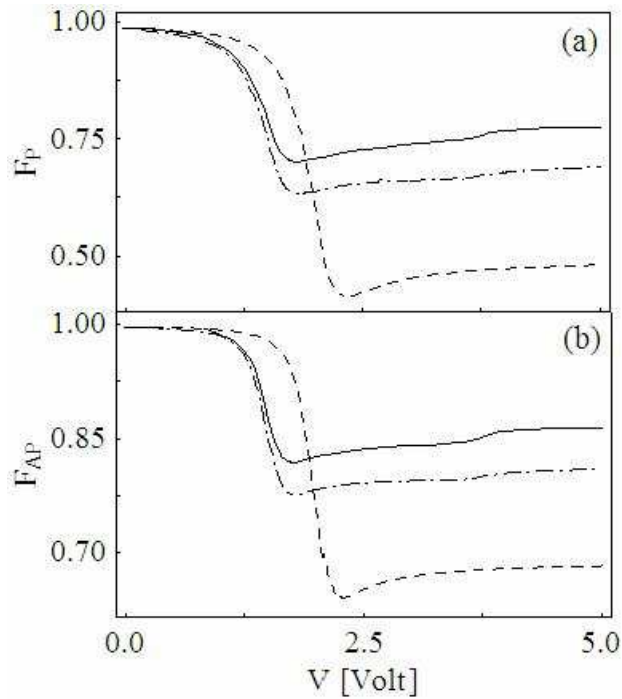

Figure 5: Fano factor $F(V) = F(-V)$ as a function of bias voltage in the case of parallel (a) and antiparallel alignment of magnetizations (b). Total Fano factor (solid line) and its elastic part (dashed-dotted line) are compared with the factor obtained in the absence of phonons (dashed line).

## 4. Concluding remarks

Summarizing, we have presented a general method that can be used to study spin-dependent electrical current and shot noise of inelastic transport through magnetic nanojunctions, using GFT-MT technique. This non-perturbative computational scheme is entirely based on mapping which transforms the many-body electron-phonon interaction problem into a single-electron multi-channel scattering problem. As an example, we have analyzed the problem of conduction through vibrating molecular bridge (quantum dot represented by one electronic state and molecular vibrations modeled as dispersionless phonons) which is connected to ferromagnetic electrodes (treated within the wide-band approximation). Our results show that transport in the presence of phonons is due to coherent propagation of polarons, where polaron shift and polaron excited states can be observed in analyzed transport characteristics. In particular, the crossover in the shot noise spectrum from Poissonian limit to sub-Poissonian region and reduction of electron correlations due to electron-phonon interaction effects are also predicted for higher voltages (after the first step in the I-V dependence).

This work brings us nearer in the direction of understanding the electrical conduction at molecular scale. However, it should be also emphasized that in the presented method we have completely ignored few important effects which can have significant influence on transport characteristics. Namely: (i) phase decoherence processes in the treatment of the electron-phonon exchange, (ii) Coulomb interactions between charge carriers and (iii) phonon mediated electron-electron interaction (i.e. formation of Cooper pairs at low temperatures). Our analysis is also based on the assumption of spin-conserving character of transport, where spin-flip scattering and all the spin-orbit processes are neglected. Such simplification can be justified by the fact that spin orientation of conduction electrons survives for a long period of time ($\sim ns$) in comparison with the residence time of the tunneling electron on the molecular bridges ($\sim fs$). Thus molecular junctions may be useful in applications involving electron spin manipulations.

Concluding, molecular junctions are important both from a pure science viewpoint and because of their potential applications. They are promising candidates for future electronic devices because of: (i) their small sizes, (ii) quantum nature of electrical conduction and (iii) theoretically inexhaustible possibilities of structural modifications of the molecules. They also have potential to become relatively cheap and easy in obtaining layer-based molecular junctions (due to self-assembly features of organic molecules). Among the most important tasks in molecular electronics we can enumerate: (i) fabrication of molecular junctions, (ii) understanding of the mechanisms of the current flowing through such devices, (iii) determination of the main factors that control transport phenomena in molecular systems, and eventually (iv) the connection of individual devices into a properly working integrated circuit (nanoIC). The final goal of molecular electronics might be construction of supercomputer with molecular processor that could have extraordinary parameters [45].